\documentclass[amsmath,amssymb, 12pt]{revtex4}

\usepackage{graphicx}
\usepackage{bm}

\begin{document}
\newcommand{\gmn}{ g_{\mu \nu} }
\newcommand{\beq}{\begin{equation}}
\newcommand{\eeq}{\end{equation}}
\newcommand{\ds}{\displaystyle}
\newcommand{\ddd}{\Phi_{,{\rho}}\Phi^{,{\rho}}}
\newcommand{\re}{{\frac{1}{2}g_{\mu\nu}}R}
\newcommand{\trr}{\Phi_{,{\mu}}\Phi_{,{\nu}}}
\newcommand{\be}{\begin{equation}}
\newcommand{\en}{\end{equation}}
\newcommand{\bea}{\begin{eqnarray}}
\newcommand{\ena}{\end{eqnarray}}
\newcommand{\des}{\baselineskip 24pt}
\topmargin -0.5cm

\title{An accelerated closed universe}
\author{Sergio del Campo$^1 \thanks{E-Mail: sdelcamp@ucv.cl}$,
Mauricio Cataldo$^2$\thanks{E-Mail: mcataldo@ubiobio.cl} and
Francisco Pe\~na$^3 \thanks{E-Mail: fpena@gacg.cl}$
  }
  \address{$^1$ Instituto de F\'\i sica, Pontificia Universidad Cat\'olica de Valpara\'\i so, \\
Av Brasil 2950, Valpara\'\i so, Chile.}
\address{$^2$ Depto. de F\'\i sica, Facultad de Ciencias, Universidad del B\'{\i}o B\'{\i}o, \\
Av. Collao 1202, Casilla 5-C, Concepci\'on, Chile.}
\address{$^3$ Departamento de Ciencias F\'\i sicas, Facultad de
Ingenier\'\i a, Ciencias y Administraci\'on, Universidad de la
Frontera, Avda. Francisco Salazar 01145, Casilla 54-D, Temuco,
Chile.\\}


\begin{abstract}
We study a model in which a closed universe with  dust and
quintessence matter components may look like an accelerated flat
Friedmann-Robertson-Walker (FRW) universe at low redshifts.
Several quantities relevant to the model are expressed in terms of
observed density parameters, $\Omega_M$ and $\Omega_{\Lambda}$,
and of the associated density parameter $\Omega_Q$ related to the
quintessence scalar field $Q$. \vspace{0.5cm}

\noindent {\it {}}
\end{abstract}


\maketitle

\section{\bf Introduction}

We still do not know the geometry of the universe. This question
is intimately related to the amount of matter present in the
universe. Observational evidences tell us that the measured matter
density of baryonic and nonbaryonic components is less than one,
i.e. its critical value. However, recent measurements of a type Ia
distant supernova (SNe Ia )\cite{Pe-etal,Ga-etal}, at redshift $z
\sim 1$, indicate that in the universe there exists an important
matter component that, in its most simple description, has the
characteristic of the cosmological constant, i.e. a vacuum energy
density which contributes to a large component of negative
pressure, and thus accelerates rather than decelerates the
expansion of the universe. Other possible interpretations have
been given for describing the astronomical data related to the
accelerating expansion of the universe. We distinguish those
related to topological defects\cite{Vi} from those which have
received a great deal of attention today: the quintessence matter
represented by a scalar field $Q$\cite{CaDaSt}.

Various tests of the cosmological standard model, including
spacetime geometry, peculiar galaxy velocities, structure
formation, and very early universe descriptions (related to
inflation and cosmic microwave background radiation), support a
flat universe scenario. Specifically, the mentioned
redshift-distance relation for supernova of type
Ia\cite{Pe-etal,Ga-etal}, anisotropies in the cosmic microwave
background radiation\cite{Me-etal} and gravitational
lensing\cite{Me}, all of them suggest that\cite{DB-etal}
 \be \ds
\Omega_M + \Omega_{\Lambda} = 1.03^{+ 0.05}_{-0.04}   , \label{O1}
\en in which $\ds \Omega_M = \left ( \frac{8 \pi G}{3 H_0^2}
\right )\,\rho_M^0 $ and $\ds
\Omega_{\Lambda}\,=\,\frac{\Lambda}{3\,H_0^2}$, where $\rho_M^0$
and $H_0$ are the present values of the matter density and the
Hubble parameter respectively. Here, $\Omega_{\Lambda}$ is the
fraction of the critical energy density contained in a smoothly
distributed vacuum energy referred to as a cosmological constant
$\Lambda$, and $\Omega_M$ represents the matter density related to
the baryonic and nobaryonic Cold Dark Matter (CDM) density. The
constant $G$ represents the Newton constant and we have taken
$c=1$. In the following, all quantities that are evaluated at
present time, i.e. at $t=t_0$, will be denoted by the subscript
$0$. Also, we will keep the value $c=1$ for the speed of light
throughout this paper.

On the light of these results an interesting question to ask is
whether this flatness may be due to a sort of compensation among
different components that enter into the dynamical equations. In
this respect, our main goal in this paper is to address this sort
of question by considering a simple model. In the literature we
find some descriptions along these lines. For instance, a closed
model has been studied with an important matter component whose
equation of state is given by $p = - \rho / 3$. Here, the universe
expands at a constant speed\cite{Ko}. Other authors, while using
the same astronomically observed properties for the universe, have
added a nonrelativistic matter density in which the total matter
density $\Omega_0$ is less than one, thus describing an open
universe\cite{KaTo}. Also, a flat decelerating universe model has
been simulated\cite{CrdCHe}. The common fact to all of these
models is that, even though the starting geometry were other than
that corresponding to the critical geometry, i.e. a flat geometry,
all these scenarios are, at low redshift, indistinguishable from a
flat geometry, and none of them have included a cosmological
constant.

In this paper we wish to consider a closed universe model composed
of two matter components: one related to the usual dust matter and
the other to quintessence matter. The geometry, together with
these matter components, confabulates in such a special way that
it gives rise to a flat accelerating universe scenario. The
parameters of the resulting model could be fixed by using
astronomical observations. Here the quintessence component is
characterized by a scalar field $Q$ that satisfies the following
equation of state:
\be
P_{_Q} = w_{_Q} \rho_{_Q}, \label{es1}
\en
where, in general, the equation of state parameter $w_{_Q}$ is
assumed to be a time dependent quantity, and its present value
runs in the range $-1 < w_{_Q} < -1/3$. At this point we note that
this range agrees with the values given to this parameter in the
original quintessence model\cite{CaDaSt}. On the other hand, from
the associated astronomical observations (related to the SNe Ia),
consistence is obtained if the equation of state parameter
$w_{_Q}$ satisfies the bound $w_{_Q} < -0.6$ at the present
time\cite{Ga-etal}. However, new measurements may decrease the
mathematical and/or statistical errors and thus this upper bound
may increase (or decrease). In any case, in this paper we shall
keep the range specified above, i.e., $-1 < w_{_Q} < -1/3$.

When the  scalar field $Q$ component is added to the relevant
baryons ( assumed to be described by $\Omega_b = 0.04 \pm 0.01$ )
and cold dark matter (contributing to the total mass  with
$\Omega_{CDM} = 0.30 \pm 0.10$) components, all added together
these give a value for $\Omega_M = 0.34 \pm 0.11$ \cite{Wa-etal},
and the resulting scenario is called a QCDM model, different from
$\Lambda$CDM where, in place of the quintessence scalar field $Q$,
is placed the cosmological constant $\Lambda$.

One of the goals that we have in mind in the present paper is to
investigate the conditions under which a scenario with positive
curvature may mimic an accelerating flat universe at low
redshifts. This idea allows us to determinate the exact
contribution of the scalar field $Q$ (together with the curvature
term) that gives rise to an effective cosmological constant term.
In this way we obtain an effective cosmological scenario that
coincides with the accepted $\Lambda$CDM model.

\section{\bf {The field equations}}

In order to write the corresponding field equations we use the
following effective Einstein action: \be \hspace{-4.0cm} \ds
S\,=\,\int{d^{4}x\,\sqrt{-g}}\,\left [\,\frac{1}{16\pi\,G}\,R\,
 +\,\frac{1}{2}\,(\partial_{\mu}Q)^2\, -\,V(Q)\,+\,L_{M}
\right ]. \label{ac1} \en Here, $R$ is the scalar curvature,
$V(Q)$ is the scalar potential associated with the quintessence
scalar field and $L_{M}$ represents the matter components other
than the $Q$-component.

Let us considering a FRW metric \be \ds d{s}^{2}= d{t}^{2}-
a(t)^{2}\, \left [ \frac{dr^2}{1-k r^2} +\,r^2 \left(\,d\theta^2+
sin^2 \theta \,d\phi^2 \right) \frac{}{}\,\right ], \label{me1}
\end{equation}
where $a(t)$ represents the scale factor, and the curvature
parameter $k$ takes the values $k\,=\,-\,1,\, 0,\, 1$
corresponding to an open, flat and closed three-geometry,
respectively. We shall assume that the $Q$ field is homogeneous,
i.e. is only a time-dependent quantity, and the geometry of the
universe is taken to be closed, i.e. $k = 1$. With these
assumptions we obtain from the action (\ref{ac1}) the following
Einstein field equations:
\be
\ds H\,^{2}\,=\,\frac{8\pi\,G}{3}\, \left
(\,\rho_{_{M}}\,+\,\rho_{_{Q}}\,\right )\, -\,\frac{1}{a^{2}},
\label{h1} \en \be \label{dh1} \dot{H}+H^2=-\frac{4 \pi G}{3}
(\rho_{_{M}} + \rho_{_{Q}}+ 3 P_{_{M}}+3P_{_{Q}}),
\en
and the evolution equation for the scalar field $Q$: \be \ds
\ddot{Q}\, +\,3\,H\,\dot{Q}\,=-
\,\frac{\partial{V(Q)}}{\partial{Q}}. \label{ddq} \en

Here the overdots stand for derivatives with respect to the time
${t}$, $ \ds H\,=\,\frac{\dot{a}}{a}$ defines the Hubble expansion
rate, $\rho_{_{M}}$ is the effective matter energy density, and
$P_{_{M}}$ is the pressure associated with this matter.
$\rho_{_{Q}}$ and $P_{_{Q}}$ are the average energy density and
average pressure related to $Q$ which we define to be given by
$\ds \rho_{_{Q}}\,=\,\frac{1}{2}\dot{Q}^2\,+\,V(Q)$ and $\ds
P_{_{Q}}\,=\,\frac{1}{2}\dot{Q}^2\,-\,V(Q)$.
These two quantities are related by the equation of state, eq.
(\ref{es1}).

This set of equations reveals a combination of two non-interacting
matter components that are represented by perfect fluids.

In the next section we study the consequences that arise when the
basic set of equations together with the corresponding equations
of states for the matter and the scalar field components (which
relate $p_{_M}$ to $\rho_{_M}$ and $p_{_Q}$ to $\rho_{_Q}$,
respectively), are used for describing a model which resembles the
standard $\Lambda$CDM model.

\section{Characteristics of the model}

In this section we shall impose conditions under which closed
universes may look similar to a flat universe at low redshift.

We start by considering a closed FRW model which has two matter
components. One of these components is related to a
nonrelativistic dust (i.e. matter whose equation of state is
$P_{_M} = 0$), and the other, the quintessence component whose
equation of state is expressed by eq.(\ref{es1}). In the
following, we shall assume that the quintessence component
together with the curvature term combine in such a way that they
give rise to a cosmological constant term in a flat universe
model. In this scenario, the cold dark matter (assumed to be
described by dust), together with a cosmological constant, form
the main matter ingredients of the model. To this goal we impose
the following condition:
\be
\label{c1}\ds \frac{8 \pi
G}{3}\rho_{_{Q}}(t) -\frac{1}{a^2(t)} = \frac{\Lambda}{3},
\en
from which we could get an explicit expression for $\rho_{_Q}$ as
a function of time if the time dependence of the scale factor
$a(t)$ is known.

The constraint equation (\ref{c1}) may be written as
\begin{eqnarray}
\ds \Omega_Q \,\left(\,\frac{\rho_{_Q}(t)}{\rho_{_Q}^0}\,
\right)\,+\,\Omega_k \,\left ( \frac{a_0}{a(t)} \right
)^2\,=\,\Omega_{\Lambda},  \label{cc1}
\end{eqnarray}
where the curvature density parameter $\Omega_k$ and the
quintessence density parameter $\Omega_Q$ are defined by $\ds
\Omega_k \,=\,-\,\left (\frac{1}{a_0\,H_0} \right
)^2\,\,(\,<\,0\,)$ and $\ds \Omega_Q\, =\,\left (\,
\frac{8\,\pi\,G}{3\,H_0^2}\, \right )\,\rho_{_Q}^0\,\,(\,>\,0\,)$
respectively. When we evaluate eq. (\ref{cc1}) at the present
epoch, we get a relation among the density parameters
\begin{eqnarray}
\ds \Omega_Q \,+\,\Omega_k \,=\,\Omega_{\Lambda}.  \label{cc3}
\end{eqnarray}
Since $\Omega_k\,<\,0$, we get that $\Omega_Q$ must be greater
than $\Omega_{\Lambda}$.

Under the condition (\ref{c1}), the time-time component of
Einstein equations, eq. (\ref{h1}), becomes analogous to that of a
flat universe where the matter density $\Omega_M$ and the
cosmological constant density $\Omega_{\Lambda}$ form the main
matter components today. Thus, this equation reads
\be
\ds
H\,^{2}(t)\,=\,H_0^2\,\left [ \,\Omega_{\Lambda}\,+\, \Omega_{M}\,
\frac{ \rho_{_M}(t)}{\rho_{_M}^0}\,\right ]. \label{h2}
\en

Notice that, when this expression is evaluated at present time,
i.e. $t=t_0$, we get $\Omega_M + \Omega_{\Lambda} = 1$, which lies
within the observational range as is seen from eq. (\ref{O1}).
Therefore, we may associate with this scenario the so called
$\Lambda$CDM-model. For numerical computations we shall take
$\Omega_M = 0.3$ and $\Omega_{\Lambda}= 0.7$. The latter choice
agrees with the value of a cosmological constant which is
constrained to be $\Omega_{\Lambda}\,\leq 0.7$ by QSO lensing
surveys\cite{Koc}.

Thus our basic equations are~(\ref{ddq}) and~(\ref{h2}), together
with the constraint equations~(\ref{cc1}). Notice that this set of
equations should not be confused with the one specified by
Starobinski~\cite{Star1,Star2}. They describe dust and a scalar
field in a flat FRW model. In our case, we take dust and scalar
field in a close FRW universe. Our scalar field is set up in such
away that we constraint the Einstein field equations to have a
flat form.

It is well known that eq.~(\ref{h2}) can be solve exactly for a
non-relativistic perfect fluid
(dust)~\cite{Lemaitre,Peebles,Star3}: \be \ds a(t)\,=\,a_0\,\left
( \frac{\Omega_M}{\Omega_{\Lambda}}\right
)^{1/3}\,\sinh^{2/3}\left ( \beta \, t\right ), \label{a1} \en
where $\beta=\frac{3}{2}\,\sqrt{\Omega_{\Lambda}}\,\,H_0$

This solution allows us to write explicit expressions for
$\rho_{_Q}$ and $\rho_M$: \be \ds
\rho_{_Q}(t)\,=\,\frac{\rho_{_Q}^0}{\Omega_Q}\,\left [
\Omega_{\Lambda}\,+\,
(\Omega_Q\,-\,\Omega_{\Lambda})\,\left(\frac{\Omega_{\Lambda}}{\Omega_M}
\right )^{2/3}  \,\sinh^{-4/3}\left (
\frac{3}{2}\,\sqrt{\Omega_{\Lambda}} \,\,H_0\,t\,\right )  \right
]
\en
and
\begin{eqnarray}
\ds \label{rm3} \rho_M(t)\,=\,\rho_M^0\,
\left(\,\frac{\Omega_{\Lambda}}{\Omega_M}\,\right
)\,\sinh^{-2}\,\left ( \frac{3}{2}\,\sqrt{\Omega_{\Lambda}}
\,\,H_0\,t \right ).
\end{eqnarray}

We observe that the energy density $\rho_M$ becomes dominant for
$t \ll t_{eq}$, where $t_{eq}$ is the time when the two fluids,
the quintessence scalar field $Q$ and the CDM components, become
identical. This time is given by \be \ds
t_{eq}\,=\,\frac{2}{3\,H_0\,\sqrt{\Omega_{\Lambda}}}
\,\sinh^{-1}\,\left \{\,\frac{\left [
\,(\,1\,+\,\sqrt{1\,+\,D})^{2/3}\,-\,D^{1/3}\,\right
]^{3/2}}{\sqrt{2\,(\,1\,+\,\sqrt{1\,+\,D}\,)}}\,\right \},
\label{t1}
\en
where the parameter $D$ is defined by $\ds
D\,=\,\frac{4}{27}\,\frac{(\Omega_Q
-\Omega_{\Lambda})^3}{\Omega_{\Lambda}\,\Omega_M^2}$.

Since we have fixed the values of $\Omega_M$ and $\Omega_\Lambda$
with present astronomical observations, the time $t_{eq}$ will
depend on the free parameter $\Omega_Q$. For instance, if we take
$\Omega_Q=0.75$, this time is quite close to the present time. In
fact, in this case, $t_{eq}$ represents approximately $99 \%$ of
$t_0$. For $\Omega_Q = 0.95$, this corresponds to $78$ percent
approximately. For the period $t > t_{eq}$ it is the scalar field
$Q$ that dominates. In particular, at present time we find
$\rho_{_Q}^0\,>\,\rho_M^0$ or, equivalently, $\Omega_Q > \Omega_M
$. And, due to the expansion acceleration measured for the
universe which implies that $\Omega_{\lambda}
> \Omega_M$, our model satisfies the following inequalities for the density
parameters:  \, $\Omega_Q > \Omega_\Lambda > \Omega_M$.

Continuing with our analysis, we would like to obtain the explicit
time dependence for the equation of state parameter $w_{_Q} $. In
order to do so, we take into account eq.~(\ref{es1}) 
and from the constraint eq.~(\ref{cc1}) we obtain that
\begin{eqnarray}
\label{wt1} \ds w_{_Q}(t)\,=\,-\,\frac{1}{3}\,\left [\frac{
\,(\Omega_Q\,-\,\Omega_\Lambda)\,\left (a_0/a(t) \right
)^2\,+\,3\,\Omega_\Lambda}{\,(\Omega_Q\,-\,\Omega_\Lambda)\,\left
(a_0/a(t) \right )^2\,+\,\Omega_\Lambda}\, \right ],
\end{eqnarray}
where $a(t)$ is given by eq. (\ref{a1}). Notice that the case
$\Lambda\,=\,0$ gives  $\ds w_{_Q}(t)\,=\,-\,1/3\,=\,$ const.,
situation that was studied in ref. \cite{CrdCHe}. For $\Lambda
\neq 0$, this parameter is always negative, since for $a
\longrightarrow 0$, the parameter $w_{_Q} \longrightarrow -1/3$
and, when $a \longrightarrow \infty$, we get $w_{_Q}
\longrightarrow -1$. Thus, we find that the parameter $w_{_Q}$
lies in the range $-1\,<\,w_{_Q}\,<\,-1/3$. We also should note
that if we would have considered the open model, i.e. the case
with $k=-1$ in the FRW metric, the equation of state parameter
would have an unattractive characteristic, since this would lie in
the range $-\infty < w_{_Q} < - 1$ violating the dominant energy
condition, i.e. $\mid P_Q \mid \leq \rho_{_q}$\cite{HaEl}.

Another interesting characteristic of the quintessence scalar
field to be determined is the form of the scalar potential,
$V(Q)$. In order to do this, we notice from definitions
$\rho_{_Q}$ and $P_{_Q}$ that the scalar potential becomes given
by $V(Q)=(1/2)(1-w_{_Q})\rho_{_Q}$ where we have used the equation
of state (\ref{es1}). If in this expression we substitute the
corresponding expressions  (\ref{cc1}) and (\ref{wt1}), we obtain
that

\be \ds V(t)\,=\,\frac{1}{3}\,\rho_{_Q}^0\,\left [\,3\,\left (
\frac{\Omega_\Lambda}{\Omega_Q} \right )\,  + \,2\,\left
(\,1\,-\,\frac{\Omega_\Lambda}{\Omega_Q}\,\right )\,\left (
\frac{a_0}{a(t)}\,\right )^2 \,\right ]. \label{vt1}
\end{equation}

On the other hand, from the same definitions for $\rho_{_Q}$ and
$P_{_Q}$ 
we get that $\dot{Q}\,=\,\sqrt{\rho_{_Q}\,\left [1\,+\,w_{_Q}\,
\right ]}$ and, after substituting the corresponding expression
for $\rho_{_Q}$ and $w_{_Q}$ , we get an explicit equation for
$\dot{Q}$, as a function of the scale factor, which can be
integrated and obtain \be \ds Q(t)\,=\frac{3
\alpha}{\beta}\,\left(\frac{\Omega_\Lambda}{\Omega_M}\right)^{1/2}\,
 \left (\frac{a(t)}{a_0}\right )^{(1/2)}  \,_2{F}_1\left
(\frac{1}{2},\frac{1}{6};\frac{7}{6};-\left (
\frac{\Omega_\Lambda}{\Omega_M}\right )\left (
\frac{a(t)}{a_0}\right)^3 \right ),\label{qa1}
\end{equation}
where $_2F_1$ is the generalized hypergeometric function and
$\alpha=\sqrt{\frac{2}{3} \, \rho^0_Q \, \left(
1-\frac{\Omega_{\Lambda}}{\Omega_{Q}} \right)}$. This expression
has been obtained by using MAPLE.

As we have mentioned below eq.~(\ref{cc3}) the range
$\Omega_{_{\Lambda}} < \Omega_{_{Q}}$ has to be satisfied. On the
other hand, an upper limit for $\Omega_{_{Q}}$ could be restricted
considering the data specified in ref. \cite{Lange}, where the
range for $\Omega_k$ is established, i.e. $ - 0.15 < \Omega_k <
-0.02$, in which case we obtain $\Omega_{_Q} < 0.75 - 0.85$. Thus,
considering that $\Omega_{{\Lambda}}$ is in the range
$\Omega_{_{\Lambda}} \sim 0.6 - 0.7$ we may write for $\Omega_{Q}$
the following range: $0.6-0.7 < \Omega_{_Q} < 0.75-0.85$.

\section{The comoving volume element}

In this section we would like to obtain some observational
consequences for our model. One important issue in astronomy is
that referred to the number count-redshift relation. The number of
galaxies in a comoving element in a solid angular area $d\Omega$
with redshift between $z$ and $z + dz$ is sensitive to the
comoving volume element $dV_C$ which we define as following for
the closed FRW metric $\ds dV_C\,=\,a_0^3\,\left (
\frac{r^2}{\sqrt{1-r^2}}\right )\,dr\,d\Omega$. This expression
gives rise to the comoving volume element as\be \ds
\frac{d\,V_C}{dz\,d\Omega}\,=\,\frac{D_m^2(z)}{\sqrt{1\,+\,\Omega_k\,H_0^2\,
D_m^2(z)}}\,\frac{dD_m(z)}{dz}, \label{dv2} \en where the proper
motion distance $D_m\,=\,a_0\,r$ was introduced.

In order to get an explicit expression for the observable comoving
volume element per solid angle and per redshift interval, we need
$D_m$ as a function of the redshift $z$. Using
$D_m(z)=D_L(z)/(1+z)$, where $D_L$ represent the luminosity
distance defined by $\ds
D_L(z)\,=\,\left(\frac{\cal{L}}{4\,\pi\,\cal{F}}\right)^{1/2}$,
where $\cal{L}$ is the rest-frame luminosity and $\cal{F}$ is an
apparent flux, we get that\cite{Star3} \be
D_m(z)\,=\,\frac{1}{H_0}\,\frac{1}{\sqrt{\Omega_Q\,-\,\Omega_\Lambda}}
\sin\,\left[\,\sqrt{\Omega_Q\,-\,\Omega_\Lambda}\,\int_0^z\,
{\frac{dx}{\sqrt{\Omega_\Lambda\,+\,\Omega_M\,(1+x)^3}}}\,\right].
\label{dm} \en With this latter expression we obtain for the
comoving volume element\be \ds
\frac{dV_C}{dz\,d\Omega}\,=\,\frac{1}{H_0^3}\,
\frac{\left[H_0\,D_m(z)\right]^2}
{\sqrt{\Omega_\Lambda\,+\,\Omega_M\,(1+z)^3\,}}
\,\sqrt{\frac{1\,-\,(\Omega_Q\,-\,\Omega_\Lambda\,)\,\left[\,H_0\,D_m(z)\,
\right]^2}{1\,+\,(\Omega_Q\,-\,\Omega_\Lambda\,)\,\left[\,H_0\,D_m(z)\,
\right]^2}}. \label{dv3}\en

Expression (\ref{dv3}) has the consequence that, given a
population of objects of constant density and determinable
distance measures, we can in principle constrain the value of the
$\Omega_Q$ parameter and determine whether the universe may be
considered to be closed, even though it looks quite flat at low
redshift. At low redshift ($z < 1$) all the models become
indistinguishable one to another. This could be seen from the fact
that at low redshift we may expand eq. (\ref{dv3}) and obtain, by
keeping the first term of the expansion as a leader term \be \ds
\frac{dV_C}{dz\,d\Omega}\,\sim \, \frac{1}{H_0^3}\,z^2, \en which
becomes completely $\Omega$-parameters independent. But, at high
enough redshift, i.e. $z > 1$, the flat model shows a
volume-per-redshift larger than those related to closed models.

\section{Conclusions}

We have studied a closed universe model in which, apart from the
usual CDM component, we have included a quintessence scalar field.
At low redshift it looks flat. This means that we have fine tuned
the quintessence component together with the curvature term for
getting a flat model in which an effective cosmological constant
$\Lambda$ is the main matter component in agreement with the
observed acceleration of the universe.

We have found the intrinsic properties of the model and
especially, the characteristics of the quintessence scalar field
$Q$. For instance, the scalar potential $V(Q)$ appears to follow
an almost inverse power-law expression coincident with those
usually evoked in models where quintessence has been taken into
account. In a similar way, another property of this field is
obtained when we impose an equation of state of the form
$P_{_Q}\,=\,w_{_Q}\,\rho_{_Q}$ where, in agreement with SNe Ia
astronomical measures, it is found that this parameter lies in the
range $-1 < w_{_Q} < -1/3$. However, we should notice that this
range appears as a condition for mimicing a flat model, and not as
an imposition coming from observational constraints.

Finally, we have described a kinematical property for our model.
Specifically, we have determined in the last section the comoving
volume per solid angle per redshift interval as a function of the
redshift. We have found that the different models (including the
flat one) become indistinguishable al low enough redshift ($z \ll
1$). However, at high redshift, i.e. at $z \geq 1$, we have found
that the closed models become distinguishable from the flat model.
Perhaps the coming astronomical programs will decide which of the
considered models describes more properly the dynamics of our
observed universe.

\section{\bf Acknowledgements}
This work is dedicated to Alberto Garc\'{\i}a's 60$^{th}$
birthday. SdC and MC were supported by COMISION NACIONAL DE
CIENCIAS Y TECNOLOGIA through Grants FONDECYT N$^0$ 1030469 and
N$^0$ 1010485, respectively. Also, SdC  was supported by PUCV-DI
grant N$^0$ 123.764/03, MC by Direcci\'{o}n de Investigaci\'{o}n
de la Universidad del B\'{\i}o-B\'{\i}o and FP was supported from
DIUFRO N$^0$ 20228.


\end{document}